\documentclass[prl,twocolumn,showpacs,amsmath,amssymb]{revtex4}

\usepackage{graphicx}
\usepackage{dcolumn}
\usepackage{bm}

\begin{document}

\title{Second generation of vortex-antivortex states in mesoscopic superconductors: stabilization by artificial pinning }
\author{R. Geurts}
\author{M. V. Milo\v{s}evi\'{c}}
\author{F. M. Peeters}
\email{francois.peeters@ua.ac.be}

\address{Departement Fysica, Universiteit Antwerpen,
Groenenborgerlaan 171, B-2020 Antwerpen, Belgium}

\date{\today}

\begin{abstract}
Antagonistic symmetries of superconducting polygons and their
induced multi-vortex states in a homogeneous magnetic field may lead
to appearance of {\it antivortices} in the vicinity of the
superconducting/normal state boundary (where mesoscopic confinement
is particularly strong). Resulting vortex-antivortex (V-Av)
molecules match the sample symmetry, but are extremely sensitive to
defects and fluctuations and remain undetected experimentally. Here
we show that V-Av states can re-appear {\it deep in the
superconducting state} due to an array of perforations in a
polygonal setting, surrounding a central hole. Such states are no
longer caused by the symmetry of the sample but rather by pinning
itself, which prevents the vortex-antivortex annihilation. As a
result, even {\it micron-size, clearly spaced V-Av molecules} can be
stabilized in large mesoscopic samples.
\end{abstract}


\maketitle

Vortices in small superconducting elements have been the focus of
scientific research ever since it was shown that a strongly confined
superconducting condensate shows enhanced critical properties
\cite{geim,mosh}. Usually individual vortices pierce the sample in
the direction of applied magnetic field, minimizing the energy of
the stray magnetic field, as well as their mutual interaction. Such
formed vortex clusters (called `multi-vortex' \cite{schw}) tend to
mimic the symmetry of the sample, due to the repulsion with the
surrounding screening (Meissner) currents. In samples of the order
of the coherence length $\xi$ and/or penetration depth $\lambda$,
the edge currents may even compress the vortex lines into a single
bundle, often referred to as a `giant-vortex' \cite{schw,kanda}.
However, in addition to these simple arguments, the effects of
lateral confinement in mesoscopic elements can have more elaborate
manifestations. One such is the appearance of {\it
vortex-antivortex} molecules, of exact the same symmetry as the host
sample.

Almost a decade since the original prediction of Chibotaru {\it et
al.} \cite{chib} the symmetry-induced antivortex remains one of the
most exciting phenomena in vortex matter in submicron
superconductors. The stability of an antivortex in {\it opposite}
magnetic field is still puzzling, and not yet verified
experimentally. Latter is not surprising, taking into account that
predicted vortex-antivortex molecules are very small in size
($\sim\xi$, somewhat larger in type-I samples \cite{slava}) and
extremely sensitive to defects \cite{meln}. Actually, in most
instances, they are simply unstable \cite{ben,bonca}.

In last several years, two methods were proposed to improve the
stability and the observability of the vortex-antivortex states in
mesoscopic polygons. In one, the magnetic field profile was altered
by a magnetic dot placed on top of the superconductor \cite{carba}.
The bipolar field of the added magnet favors the antivortex
underneath, and repels vortices further apart. However, this method:
(i) interferes with the key concept of an antivortex in opposite,
unipolar field, (ii) makes the structure more complicated,
three-dimensional, and inaccessible for scanning probe techniques,
and (iii) poses difficulties for any magnetic (magnetic-force,
Hall-probe) measurements. Instead, we recently proposed the
structural engineering of the sample itself, by strategically placed
holes in the sample, mimicking the sample symmetry and the expected
symmetry of the vortex-antivortex state \cite{roel}. In such
realization each hole hosts (pins) one vortex, where the pinning
force is effectively stronger than the vortex-antivortex attraction.
As a consequence, vortices remain in the holes even when placed at
further distances from the antivortex, and the whole molecule can be
made significantly larger.
\begin{figure}[b]
\includegraphics[width=0.65\linewidth]{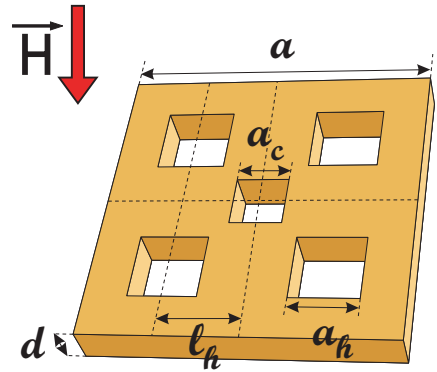}
\caption{\label{fig1} A schematic drawing of a polygonal mesoscopic
superconductor (side $a$, thickness $d$) with corresponding
arrangement of perforations (spacing $l_h$, characteristic size of
each hole $a_h$ and of the central hole $a_{c}$) in a perpendicular
magnetic field $H$.}
\end{figure}

In this paper, we realized that not only vortices can be pinned in
the latter concept, but an antivortex as well. We therefore
introduced an additional, central hole in the sample, in which the
antivortex can reside. Intuitively, this facilitates the
quantization of negative stray flux between the vortices in a
cluster, and leads to a more stable antivortex. Vortices and
antivortices in holes are just quantized magnetic field, screening
prevents their annihilation, and the fact that they have no core
reduces their attractive interaction. As will be shown, this indeed
leads to an enhanced stabilization of the vortex-antivortex
molecule, even deep into the superconducting state, but with
significantly altered overall properties. Although this generally
applies to any polygonal setting of the holes and various shapes of
the sample \cite{roelprb}, in what follows we show the proof of
principle for a square superconducting geometry, as shown in Fig.
\ref{fig1}.

\begin{figure*}[t]
\includegraphics[width=\linewidth]{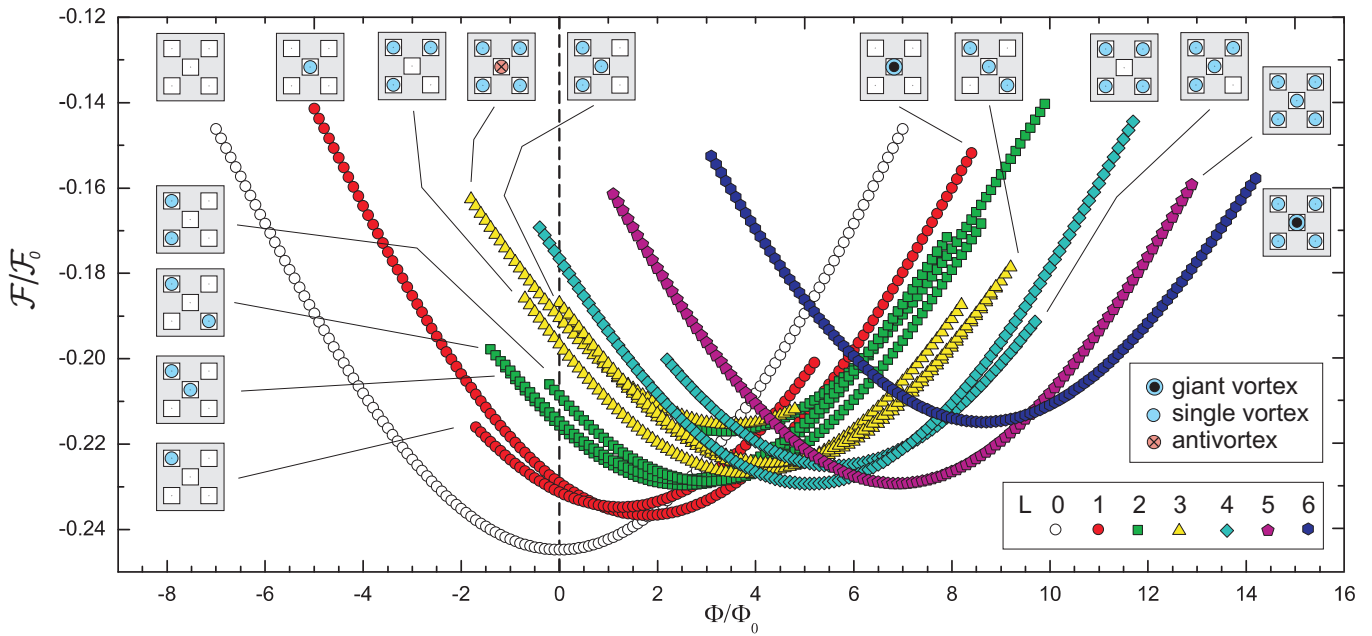}
\caption{\label{fig2} The Gibbs free energy spectrum, showing the
energy levels and stability of different vortex states as a function
of applied magnetic field at $T=0.3T_c$. Insets cartoon the vortex
arrangement for selected states.}
\end{figure*}

For the purpose of this study, we us the mean-field, Ginzburg-
Landau (GL) theory, and solve numerically the set of GL equations
\begin{eqnarray}
(-i\nabla - \vec{A})^2 \psi - (1 - T -
|\psi|^2) \psi  = 0, \label{gl1}\\
-\kappa^2 \Delta \vec{A} = \vec{j}_s  = \Re\left[ \psi\left( i\nabla
- \vec{A}\right)\psi^{*} \right]\, \label{gl2}
\end{eqnarray}
already established as a prime tool for theoretical description of
mesoscopic superconductors \cite{schw}. All variables in Eqs.
(\ref{gl1},\ref{gl2}) are dimensionless and expressed in temperature
independent units (assuming $\sim(1-T)^{-1/2}$ dependence of
characteristic length scales in the system, coherence length $\xi$,
and penetration depth $\lambda$). $\kappa=\lambda/\xi$ is the
so-called GL parameter, and describes the ability of the sample to
screen the applied magnetic field. Temperature is scaled to $T_c$,
magnetic field to $H_{c2}(0)$, vector potential to $H_{c2}(0)\xi(0)$
and the order parameter $\psi$ to $\psi(0)=\sqrt{\alpha(0)/\beta}$.
The Gibbs free energy density $\mathcal{F}$, in units of
$\mathcal{F}$(T=0, H=0)$=H_c(0)^2/8\pi$ is calculated from
\begin{eqnarray}
\mathcal{F}  = \frac{1}{V} \int & \Big[ \left| (-i\nabla - \vec{A})
\psi \right|^2 - (1-T- \frac{1}{2} |\psi|^2 ) |\psi|^2 \\
 & + \kappa^2 ( \nabla \times \vec{A} - \vec{H})^2 \Big] dV,
\label{freeen}
\end{eqnarray}
for each of the stable superconducting states found during the
simulation. The full simulation region, used for the calculation of
the magnetic response of the sample, was a square with $128$ grid
points in each direction. The number of grid points {\it inside} the
sample was kept at $64\times64$.


In Fig. \ref{fig2} we show the free energy of vortex states with
vorticities from $L=0$ to $L=6$ as a function of the applied
magnetic flux $\Phi=a^2H$. The simulation is performed by sweeping
the magnetic field up and down, and then back-tracking of all found
vortex states so that their complete stability range is obtained. In
addition, we performed the calculation starting from different
initial conditions, some of which included reasonable guesses of
possible vortex configurations. Fig. \ref{fig2} summarizes all the
found stable states, but we may not rule out the possibility of some
more complex, higher energy, vortex states. For this simulation a
square of size $20 \xi(0) \times 20 \xi(0)$ was used with holes of
size $3.5 \xi(0) \times 3.5 \xi(0)$, four of which are horizontally
and vertically displaced by $5 \xi(0)$ from the center of the
sample. Temperature was fixed at $0.3 T_c$ (corresponding to Al
samples below 400 mK \cite{khotk}) and an extreme type-II behavior
was assumed, typical for thin samples ($\kappa=\infty$). The
combinatorial number of possible vortex configurations for given
vorticity is quite high, and many of them are indeed found stable
due to both the large size of the sample and that of the central
hole. Starting from zero field, the Meissner state, $L=0$, is the
lowest energy state in the flux band of $\Delta \Phi=1.85\Phi_0$,
which is larger than a flux quantum, but significantly smaller than
in non-perforated mesoscopic samples \cite{ben}. The reason is that
the first vortex penetrates easier, and is more stable in the
present system due to the central hole (central $L=1$ is in the
ground state in $\Delta \Phi=1.92\Phi_0$). States with an off-center
vortex have mostly higher energy, smaller stability range, and are
fourfold degenerate (which makes them interesting for logic
applications \cite{fca}). $L=2$ shows the {\it broken symmetry} in
the ground state, as two vortices reside in the central and corner
hole, but their ground-state flux-band is only $\Delta
\Phi=0.17\Phi_0$. Similarly, the broken symmetry $L=3$ state with
all vortices in the corners is the ground state in a short interval
of $\Delta \Phi=0.11\Phi_0$. Just like in the superconducting square
with $2\times2$ holes \cite{golib}, $L=4$ vortices reside in the
corners and show enhanced stability in the ground state ($\Delta
\Phi=1.95\Phi_0$). However, the most dominant state in Fig.
\ref{fig2} is the $L=5$ one, where all holes are occupied by a
single vortex ($\Delta \Phi=3.63\Phi_0$). Therefore, while for the
square with an $2\times 2$ holes high symmetry states as $L=2$ and
$L=4$ were pronouncedly low-energy states, we notice that in the
present system the states which have a vortex in the central hole,
and are fourfold symmetric dominate the ground state, i.e. $L=1$ and
$L=5$. Note that the $L=4$ state is still very comfortable with the
fourfold symmetry although it does not have a central vortex, but
e.g. $L=3$ with a central vortex is not in the ground state because
it does not obey the symmetry of the sample. Because of their very
low energy, $L=1,4,5$ states overshadow other states like the $L=2$
and $L=3$. The $L=3$ state has several allotropes, i.e. it can have
vortices across the sample diagonal, one vortex in the center and
two on the side, or all three out of the center. Only one
configuration can satisfy the four-fold symmetry of the sample and
that is the $L=4-1$ state, with four vortices in corner holes and an
{\it antivortex} in the center. This vortex-antivortex (V-Av) state
is indeed found stable in our system, with relatively high energy,
but still lower than most higher vorticity states (which suggests
that it can be experimentally realized in decreasing magnetic
field).

\begin{figure}[b]
\includegraphics[width=\linewidth]{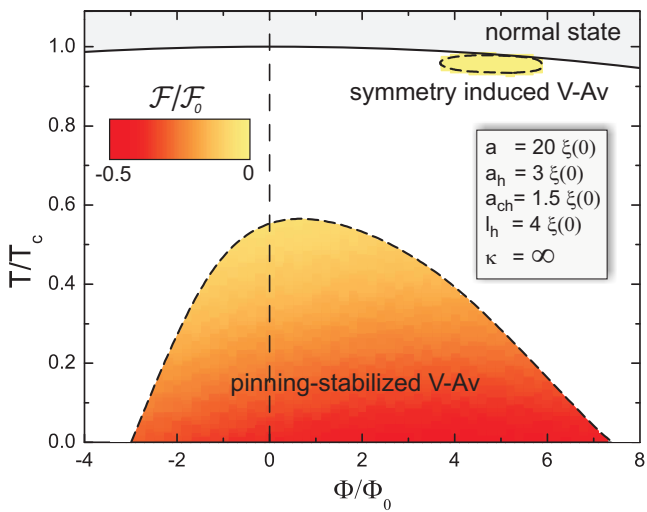}
\caption{\label{fig3} The superconducting state in the $\Phi-T$
diagram (solid line shows the superconducting/normal state boundary)
comprising the areas of stability of the $L=3$ vortex-antivortex
molecules, both pinning- and symmetry-induced ones. Color coding
shows the energy of those vortex states.}
\end{figure}

The physical origin of this V-Av state is however different from the
V-Av state found in Refs. \cite{chib,slava,meln,roel} where the V-Av
molecules appear due to the imposed symmetry of the sample. In Ref.
\cite{roel}, the added holes only reinforce the symmetry argument
and do not cause {\it per se} the V-Av state. In the rest of the
paper, we will refer to this state as the symmetry induced V-Av in
contrast to the newly found V-Av state which is {\it fully
stabilized by pinning} of all vortices and the antivortex. In Fig.
\ref{fig3} we show the $\Phi-T$ phase diagram for both versions of
the $L=3$ vortex-antivortex state (for taken parameters
$a=20\xi(0)$, $a_h=3\xi(0)$, $l_h=4\xi(0)$, $a_{c}=1.5\xi(0)$, see
Fig. \ref{fig1}). We observed two fully independent islands in the
$\Phi-T$ phase space corresponding to the two manifestations of the
V-Av state. The symmetry-induced one is situated in the high
temperature regime, where confinement imposed by the sample
boundaries is effectively very large. It should be noted here that
the $\Delta T$ stability range for the symmetry-induced V-Av state
is on average just 1\% larger than in the case of $2\times 2$ holes
\cite{roel} (since the order parameter inside the molecule is in any
case severely suppressed). At lower temperatures, the GL equations
are strongly non-linear and symmetry arguments cannot account for
the nucleation of the V-Av state. Nevertheless, a second generation
V-Av state does stabilize, thanks to the large spacing between the
holes (preventing vortex-antivortex annihilation) and the large size
of the sample (diminishing the effect of encircling screening
currents). The pinning-stabilized V-Av state: 1) occupies far larger
$\Phi-T$ area compared to the symmetry induced one, a wide region of
$\approx 10 \Phi_0$ and $\approx 0.6 T_c$, and 2) it is found stable
even {\it in negative applied field}, where central antivortex is a
natural state and vortices are subjected to an increasing expulsion
pressure. In principle, several vortex states are found stable at
negative fields in Fig. \ref{fig2}, which is a manifestation of the
flux-trapping effect \cite{fltrap}. However, out of all $L=3$
states, $L=4-1$ shows maximal resilience to negative flux.

\begin{figure}[b]
\includegraphics[width=\linewidth]{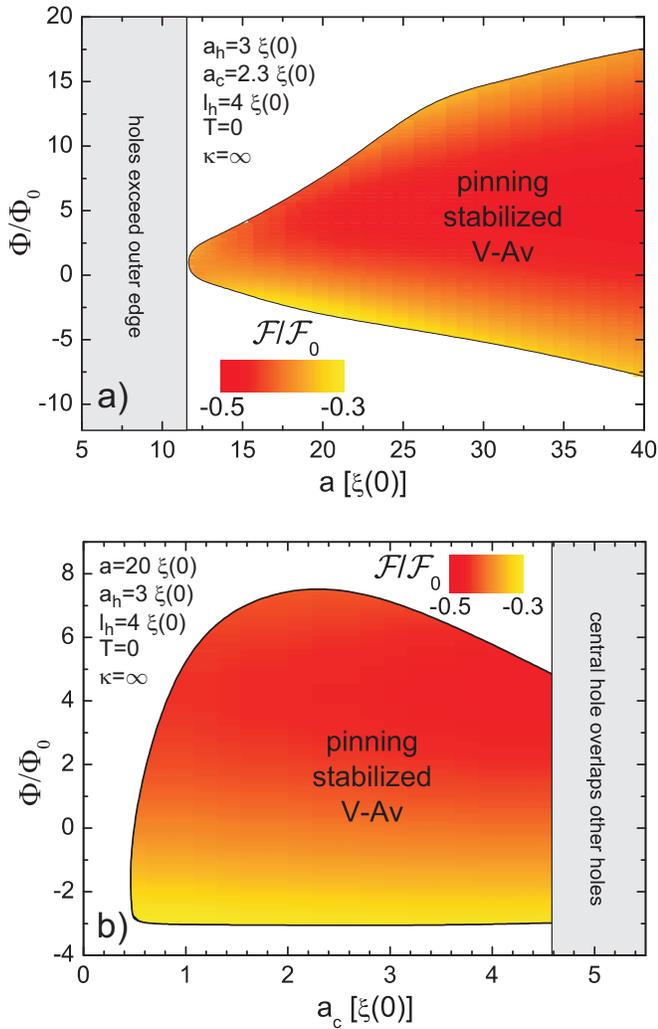}
\caption{\label{fig4} The stability diagram vs. applied field (i.e.
magnetic flux) for the pinning-induced vortex-antivortex state, (a)
as a function of the sample size for a fixed temperature, and (b) as
a function of the size of the central hole.}
\end{figure}

To emphasize again, the size of the sample is playing a crucial role
for the V-Av state. While the symmetry-induced V-Av exists only for
small samples [compared to the coherence length $\xi(T)$] the
pinning-stabilized V-Av requires a large sample. This is illustrated
in Fig. \ref{fig4}(a) for a square with five holes. When the sample
is larger, the flux stability region of the pinning stabilized V-Av
state becomes larger as well. This we can attribute to the weakened
influence of the Meissner current on the inside vortices because of
the larger distance. One also notices the extremely large flux
interval (from $-7.5 \Phi_0$ to $17 \Phi_0$) for stability of the
$L=3$ pinning-stabilized V-Av state. However, penetration field for
new vortices {\it must decrease} for larger samples (see e.g.
\cite{conn} for the case of large BiSCCO disks). It is already known
that magnetic field higher than $H_{c1}$ is needed for penetration
of vortices into mesoscopic superconductors, but this factor
decreases to unity in bulk systems. Nevertheless, latter factor
decreases with size of the sample much slower than the square power
increase of the area of the sample. As a consequence, the threshold
flux for penetration of new vortices in our sample {\it increases}
as a function of $a/\xi(0)$. We also observed the changing curvature
of the latter dependence (at $a/\xi(0)\approx 25$); from the
calculation of threshold magnetic field, we found that samples with
size of 10-25$\xi(0)$ exhibit almost identical demagnetization
effects i.e. similar vortex penetration field which then gradually
decreases for sizes above 25$\xi(0)$.

The size of the central hole is equally important as the sample
size, as it relates to the ability of V-Av pair to annihilate.
Consequently, this second generation V-Av state {\it cannot be
stabilized} in the absence of the central hole [see Fig.
\ref{fig4}(b)]. The effect of decreasing the central hole size is
also indicated in Fig. \ref{fig4}(b) - the upper boundary of the
flux stability interval is decreasing, due to facilitated V-Av
annihilation (at high fields, vortices are pushed towards the center
of the sample), while the lower boundary is in fact {\it not
influenced} by the size of the central hole. Note that lower
boundary lies in the negative field region, where vortices have
expulsion tendency, and pinning of any strength stabilizes the
central antivortex with ease.

\begin{figure}[t]
\includegraphics[width=\linewidth]{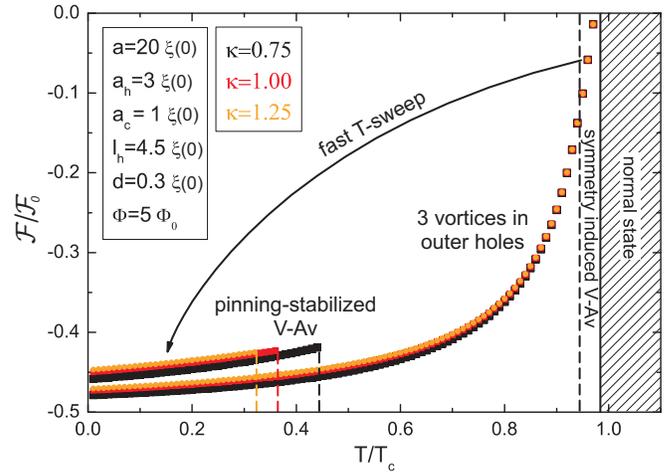}
\caption{\label{fig5} The energy of V-Av molecules as a function of
temperature, for different values of the Ginzburg-Landau parameter
$\kappa$. In numerical simulations, an abrupt cooling enables direct
transition between two kinds of V-Av states.}
\end{figure}

The latter stability of the pinned V-Av state in the negative field
forms a base for the realization of this state in experiment.
Namely, abrupt increase of the magnetic field from a low negative
value (that stabilizes one antivortex in the sample) to a large
positive one (allowing for penetration of multiple vortices in outer
holes) may result in the desired V-Av state at low temperatures.
Alternatively, following the results of Fig. \ref{fig3}, one can
think of an abrupt temperature decrease. We performed a
corresponding simulation, where we started from the symmetry-induced
V-Av state (which is the ground state for $L=3$ at temperatures near
$T_c$) and then applied a steep temperature decrease. Without taking
into account the experimentally relevant temperature gradient over
time, the simulations were always able to land into the
pinning-stabilized V-Av state at low temperatures. This is
illustrated in Fig. \ref{fig5} where also the influence of $\kappa$
is depicted. Stronger screening of the magnetic field into the holes
(i.e. lower $\kappa$) enhances the pinning, and consequently favors
the V-Av state of second generation, {\it in contrast} to the
symmetry-induced V-Av which is disfavored by low values of $\kappa$
due to the effective attraction of vortices into a single bundle
(see Ref. \cite{roel}).

\begin{figure}[t]
\includegraphics[width=\linewidth]{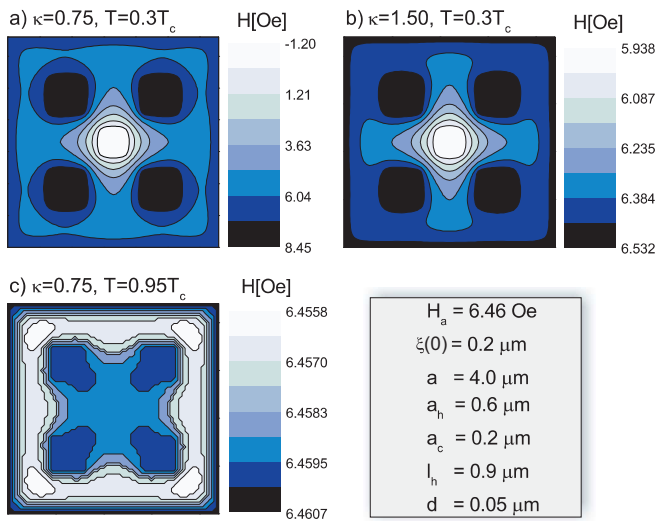}
\caption{\label{fig6} Calculated magnetic field profiles in the
sample: (a,b) for the pinning-stabilized V-Av state and two values
of $\kappa$, and (c) the corresponding profile for the
symmetry-induced V-Av state.}
\end{figure}

Although quite difficult to realize (never being the ground-state of
the system), the second generation V-Av state is much more suitable
for direct experimental visualization than its symmetry-induced
ancestor. Firstly, samples can be made larger, and spacing between
vortices and the antivortex are sufficient for their identification.
Second, the Cooper-pair density is significantly higher inside the
V-Av molecule, since temperature may be far below $T_c$. Third, the
magnetic field profile in the sample is sufficiently inhomogeneous,
with pronounced amplitudes, benefiting from the size of the V-Av
state and its stability for low values of $\kappa$. This is shown in
Fig. \ref{fig6} where the magnetic field profile in and nearby the
sample is calculated for two values of $\kappa$ for the second
generation V-Av and for one value for the symmetry-induced V-Av
(given in real units). For taken parameters, the magnetic field
contrast inside the symmetry-induced V-Av molecule is of the order
of $0.01$ Oe, while that of the second generation V-Av can be above
$10$ Oe, thus three orders of magnitude larger.

\paragraph{Conclusion}
Using a square superconducting geometry with five holes, we found a
second generation of the $L=3$ vortex-antivortex (V-Av) state in a
uniform field. This state is purely induced by pinning, and is NOT
caused by the symmetry of the sample, contrary to the previously
found V-Av molecules. This novel state is energetically favored in
rather large mesoscopic samples, but it is never the ground state of
the system. Consequently, more elaborate techniques are needed for
its stabilization in experiment, such as an abrupt increase of field
(from negative to positive value), or an abrupt cooling of the
symmetry-induced V-Av state. However, the very large $\Phi-T$
stability range of the second generation V-Av state, the very
comprehensive size of the V-Av molecule, and the large variation in
amplitudes of both superconducting order parameter and stray
magnetic field inside the molecule (further enhanced for lower
$\kappa$ and temperature), all lead to facilitated experimental
observation.

This work was supported by the Flemish Science Foundation (FWO-Vl),
the Interuniversity Attraction Poles (IAP), Programme-Belgian
State-Belgian Science Policy, ESF-VORTEX program, and the ESF-AERO
network. M.V.M. acknowledges support from the EU Marie Curie Intra-
European program.


\begin{thebibliography}{0}
\bibitem{geim} A.K. Geim, I.V. Grigorieva, S.V. Dubonos, J.G.S. Lok, J.C. Maan, A.E. Filippov,
and F.M. Peeters, Nature (London) {\bf 390}, 259 (1997); A.K. Geim,
S.V. Dubonos, I.V. Grigorieva, K.S. Novoselov, F.M. Peeters, and
V.A. Schweigert, Nature (London) {\bf 407}, 55 (2000).
\bibitem{mosh} V.V. Moshchalkov, L. Gielen, C. Strunk, R. Jonckheere, X. Qiu, C. Van Haesendonck, and Y.
Bruynseraede, Nature (London) {\bf 373}, 319 (1995).
\bibitem{schw} V.A. Schweigert, F.M. Peeters, and P.S. Deo , Phys. Rev. Lett. {\bf 81}, 2783 (1998).
\bibitem{kanda} A. Kanda, B.J. Baelus, F.M. Peeters, K. Kadowaki, and Y. Ootuka, Phys. Rev. Lett. {\bf 93}, 257002 (2004).
\bibitem{chib} L.F. Chibotaru, A. Ceulemans, V. Bruyndoncx, and V.V. Moshchalkov, Nature (London) {\bf 408}, 833
(2000).
\bibitem{slava} V.R. Misko, V.M. Fomin, J.T. Devreese, and V.V.
Moshchalkov, Phys. Rev. Lett. {\bf 90}, 147003 (2003).
\bibitem{meln} A.S. Mel'nikov, I.M. Nefedov, D.A. Ryzhov, I.A. Shereshevskii, V.M. Vinokur, and
P.P. Vysheslavtsev, Phys. Rev. B {\bf 65}, 140503(R) (2002).
\bibitem{bonca} J. Bon\v{c}a and V.V. Kabanov, Phys. Rev. B {\bf 65}, 012509
(2001).
\bibitem{ben} B.J. Baelus and F.M. Peeters, Phys. Rev. B {\bf 65}, 104515
(2002).
\bibitem{carba} C. Carballeira, V.V. Moshchalkov, L.F. Chibotaru, and A.
Ceulemans, Phys. Rev. Lett. {\bf 95}, 237003 (2005).
\bibitem{roel} R. Geurts, M.V. Milo\v{s}evi\'{c}, and F.M. Peeters, Phys. Rev. Lett. {\bf 97}, 137002 (2006).
\bibitem{roelprb} R. Geurts, M.V. Milo\v{s}evi\'{c}, and F.M. Peeters, Phys. Rev. B {\bf 75}, 184511 (2007).
\bibitem{khotk} V.V. Khotkevych, M.V. Milo\v{s}evi\'{c}, and S.J. Bending, Rev. Sci. Instrum. {\bf 79}, 123708 (2008).
\bibitem{fca} M.V. Milo\v{s}evi\'{c}, G.R. Berdiyorov, and F.M. Peeters, Appl. Phys. Lett. {\bf 91}, 212501 (2007).
\bibitem{golib} G.R. Berdiyorov, B.J. Baelus, M.V. Milo\v{s}evi\'{c}, and F. M. Peeters, Phys. Rev. B {\bf 68}, 174521 (2003).
\bibitem{fltrap} J. Jurisson, and R. J. Oakes, Phys. Lett. {\bf 2} 187 (1962);  E. Coskun, Appl. Math. and Comp. {\bf 106}, 31-49
(1999).
\bibitem{conn} M.R. Connolly, M.V. Milo\v{s}evi\'{c}, S.J. Bending, and T. Tamegai, Phys. Rev. B {\bf 78}, 132501 (2008).
\end{thebibliography}
\end{document}